\documentclass[journal,a4paper]{IEEEtran}

\usepackage[]{graphicx}
\usepackage{url}
\usepackage{ifpdf}
\usepackage{hyperref}
\hypersetup{breaklinks,colorlinks,citecolor=blue}
\usepackage{cite}
\usepackage{amssymb}
\usepackage{amsmath}
\usepackage{mathtools}

\usepackage{subfigure}
\usepackage{color}

\newcommand{\eqn}[1]{(#1)}
\newcommand{\Eqn}[1]{(#1)}

\newcommand{\fig}[1]{Fig.~#1}

\newcommand{\sectn}[1]{Sec.~#1}

\newcommand{\etal}{\mbox{\it et al.}}
\newcommand{\eg}{\mbox{\it e.g.}}

\newcommand{\cf}{\mbox{\it cf.}}

\newcommand{\sshtcode}{{\tt SSHT}}
\newcommand{\sothreecode}{{\tt SO3}}

\newcommand{\spcend}{\ensuremath{\:}}
\newcommand{\img}{\ensuremath{{\rm i}}}
\newcommand{\cconj}{\ensuremath{\ast}}

\newcommand{\integers}{\ensuremath{\mathbb{Z}}}
\newcommand{\naturals}{\ensuremath{\mathbb{N}}}

\newcommand{\ltwo}{\ensuremath{\mathrm{L}^2}}

\newcommand{\sothree}{\ensuremath{{\mathrm{SO}(3)}}}

\newcommand{\dx}{\ensuremath{\mathrm{\,d}}}

\newcommand{\deul}[1]{\ensuremath{\dx \varrho(#1)}}

\newcommand{\innerp}[2]{\ensuremath{\langle {#1},\: {#2} \rangle}}

\newcommand{\saa}{\ensuremath{\theta}}
\newcommand{\sab}{\ensuremath{\varphi}}
\newcommand{\sas}{\ensuremath{\saa, \sab}}
\newcommand{\eul}{\ensuremath{\mathbf{\rho}}}
\newcommand{\euls}{\ensuremath{\eula, \eulb, \eulc}}
\newcommand{\eula}{\ensuremath{\alpha}}
\newcommand{\eulb}{\ensuremath{\beta}}
\newcommand{\eulc}{\ensuremath{\gamma}}
\newcommand{\eulai}{\ensuremath{a}}
\newcommand{\eulbi}{\ensuremath{b}}
\newcommand{\eulci}{\ensuremath{g}}
\newcommand{\eulaiang}{\ensuremath{\eula_\eulai}}
\newcommand{\eulbiang}{\ensuremath{\eulb_\eulbi}}
\newcommand{\eulciang}{\ensuremath{\eulc_\eulci}}
\newcommand{\el}{\ensuremath{\ell}}
\newcommand{\m}{\ensuremath{m}}
\newcommand{\n}{\ensuremath{n}}

\newcommand{\spin}{\ensuremath{s}}

\newcommand{\elmax}{\ensuremath{{L}}}
\newcommand{\mmax}{\ensuremath{{M}}}
\newcommand{\nmax}{\ensuremath{{N}}}

\newcommand{\p}{\ensuremath{^\prime}}

\newcommand{\kron}[2]{\ensuremath{\delta_{{#1}{#2}}}}

\renewcommand{\exp}[1]{\ensuremath{{\rm e}^{#1}}}

\newcommand{\sshfarg}[4]{\ensuremath{{{}_{#4} Y_{#1#2}({#3})}}}
\newcommand{\sshfargc}[4]{\ensuremath{{{}_{#4} Y_{#1#2}^\cconj({#3})}}}

\newcommand{\dmatbig}{\ensuremath{D}}
\newcommand{\Dlmn}{\ensuremath{ \dmatbig_{\m\n}^{\el} }}
\newcommand{\Dlmnc}{\ensuremath{ \dmatbig_{\m\n}^{\el\cconj} }}

\newcommand{\Dlmnp}{\ensuremath{ \dmatbig_{\m\n}^{\el}(\eul) }}
\newcommand{\Dlmnpc}{\ensuremath{ \dmatbig_{\m\n}^{\el\cconj}(\eul) }}

\newcommand{\dmatsmall}{\ensuremath{d}}
\newcommand{\dlmn}{\ensuremath{ \dmatsmall_{\m\n}^{\el} }}

\newcommand{\dlmnb}{\ensuremath{ \dmatsmall_{\m\n}^{\el}(\eulb) }}

\newcommand{\dlmnhalfpi}[3]{\ensuremath{ \Delta_{{#2}{#3}}^{#1} }}

\newcommand{\dlmnhalfpim}{\ensuremath{ \Delta_{{\m\p}{\m}}^{\el} }}
\newcommand{\dlmnhalfpin}{\ensuremath{ \Delta_{{\m\p}{\n}}^{\el} }}

\newcommand{\wigc}[4]{\ensuremath{{#1}^{#2}_{{#3}{#4}}}}

\newcommand{\f}{\ensuremath{f}}

\newcommand{\sumlmn}{\ensuremath{\sum_{\el=0}^{\infty} \sum_{\m=-\el}^\el} \sum_{\n=-\el}^\el}

\newcommand{\Gmn}{\ensuremath{G_{\m\n}}}
\newcommand{\Gmnm}{\ensuremath{G_{\m\n\m\p}}}
\newcommand{\Gmnb}{\ensuremath{G_{\m\n}(\eulb)}}

\newcommand{\rGmnb}{\ensuremath{\tilde{G}_{\m\n}(\eulb)}}

\newcommand{\Fmnm}{\ensuremath{\tilde{G}_{\m\n\m\p}}}
\newcommand{\Fmnmp}{\ensuremath{\tilde{G}_{\m\n\m\p{}\p}}}
\newcommand{\FImnm}{\ensuremath{F_{\m\n\m\p}}}

\newcommand{\summptrunc}{\ensuremath{\sum_{\m\p=-(\elmax-1)}^{\elmax-1}}}
\newcommand{\qweight}{\ensuremath{q}}

\newcommand{\weighttrans}{\ensuremath{v}}

\newcommand{\weight}{\ensuremath{w}}
\newcommand{\order}{\ensuremath{\mathcal{O}}}

\renewcommand{\eqn}[1]{Eqn.~(#1)}
\renewcommand{\Eqn}[1]{Eqn.~(#1)}

\begin{document}
\title{A novel sampling theorem on the rotation group}
\author{Jason~D.~McEwen, Martin~B\"uttner, Boris~Leistedt,
  Hiranya~V.~Peiris and Yves~Wiaux%
  \thanks{J.~D.~McEwen was supported by the Engineering and Physical
    Sciences Research Council (grant number
    EP/M011852/1). M.~B\"uttner was supported by a New Frontiers in
    Astronomy and Cosmology grant \#37426 and by the Royal
    Astronomical Society. B.~Leistedt was supported by the IMPACT and
    Perren funds. B.~Leistedt and H.~V.~Peiris were supported by the
    European Research Council under the European Community's Seventh
    Framework Programme (FP7/2007-2013) / ERC grant agreement no
    306478-CosmicDawn.}
  \thanks{J.~D.~McEwen is with the Mullard Space Science Laboratory
    (MSSL), University College London (UCL), Surrey RH5 6NT, UK.
    M.~B\"uttner, B.~Leistedt and H.~V.~Peiris are with the Department
    of Physics and Astronomy, UCL, London
    WC1E 6BT, UK.  Y.~Wiaux is with the Institute of Sensors, Signals,
    and Systems, Heriot-Watt University, Edinburgh EH14 4AS, UK.}%
  \thanks{E-mail: jason.mcewen@ucl.ac.uk (J.~D.~McEwen)}}

\markboth{IEEE SIGNAL PROCESSING LETTERS,~Vol.~--, No.~--, --}%
{McEwen \MakeLowercase{\textit{et al.}}: A novel sampling theorem on
  the rotation group}
% \markboth{Journal of \LaTeX\ Class Files,~Vol.~11, No.~4, December~2012}%
% {Shell \MakeLowercase{\textit{et al.}}: Bare Demo of IEEEtran.cls for Journals}



\maketitle

\begin{abstract}
  We develop a novel sampling theorem for functions defined on the
  three-dimensional rotation group \sothree\ by connecting the
  rotation group to the three-torus through a periodic extension.
  Our sampling theorem requires $4\elmax^3$ samples to capture all of
  the information content of a signal band-limited at $\elmax$,
  reducing the number of required samples by a factor of two compared
  to other equiangular sampling theorems.  We present fast algorithms
  to compute the associated Fourier transform on the rotation group,
  the so-called Wigner transform, which scale as $\order(\elmax^4)$,
  compared to the naive scaling of $\order(\elmax^6)$. For the common
  case of a low directional band-limit $\nmax$, complexity is reduced
  to $\order(\nmax \elmax^3)$.  Our fast algorithms will be of direct
  use in speeding up the computation of directional wavelet transforms
  on the sphere.  We make our \sothreecode\ code implementing these
  algorithms publicly available.
\end{abstract}

%==============================================================================

% \fi
\IEEEpeerreviewmaketitle

%==============================================================================
\section{Introduction}

\IEEEPARstart{S}{hannon} established the theoretical foundations of
sampling theory in Euclidean space over half a century ago, proving
that the information content of a band-limited continuous signal
could be captured completely in a finite number of samples \cite{shannon:1949}.
The fast Fourier transform (FFT) \cite{cooley:1965} is one of the most
important algorithmic developments of our era and has been
instrumental in rendering the frequency content of signals accessible
in practice.  The combination of theoretical foundations and fast
algorithms has led to the extensive use of Fourier methods to analyse
data in myriad applications. % throughout science and engineering.

In many applications, however, data are acquired on non-Euclidean
manifolds where Euclidean sampling theory is not applicable.  
% The dearth of mature theory and fast algorithms in these domains
% limits the applicability of Fourier analysis techniques severely.
Spherical manifolds are one of the most prevalent non-Euclidean
domains. When observing over directions,
data are acquired on the two-dimensional sphere.  If distance
information is also accessible, then data are acquired on the
three-dimensional ball.  For example, in cosmology observations are
made on the celestial sphere, with the cosmic microwave background (CMB), observed on
the sphere (\eg\ \cite{planck2013-p01}), while surveys of the distribution of
galaxies are made on the ball (\eg\ \cite{ahn:2012}).  
% Such data-sets can be very large (\eg\ CMB data observed by the Planck
% satellite \cite{planck2013-p01}).
Other examples of data defined on spherical manifolds are found in
many fields, including 
%computer graphics (\eg\ \cite{ramamoorthi:2004}),
planetary science (\eg\ \cite{audet:2010,audet:2014}), 
geophysics (\eg\ \cite{swenson:2002}) 
and biomedical imaging (\eg\ \cite{tuch:2004}).
% planetary science (\eg\ \cite{audet:2010,audet:2014}), 
% geophysics (\eg\ \cite{simons:2006,swenson:2002}) 
% and \revised{biomedical imaging (\eg\ \cite{tuch:2004,descoteaux:2010})}.

Sampling theory and corresponding fast harmonic transforms on the
sphere remain active areas of research.  Until recently, the canonical
equiangular sampling theorem on the sphere was that of Driscoll \&
Healy \cite{driscoll:1994}, requiring $\sim 4 \elmax^2$ samples on the
sphere to capture the information content of a signal band-limited at
$\elmax$. A novel sampling theorem on the sphere was developed
recently by two of the authors of the present article (McEwen \&
Wiaux), reducing the number of samples required to capture a
band-limited signal to $\sim 2 \elmax^2$ \cite{mcewen:fssht}, building
on the developments of \cite{mcewen:fsht, huffenberger:2010} (see
\cite{mcewen:fssht} for a review of sampling theory on the sphere).
No existing sampling theorem on the sphere reaches the optimal number of
samples given by the harmonic dimensionality of a band-limited signal
of $\elmax^2$. A new sampling scheme that achieves the optimal number
of $\elmax^2$ samples was developed recently by
\cite{khalid:optimal_sampling}.  Whilst this scheme does not lead to a
sampling theorem with theoretically exact spherical harmonic
transforms, good numerical accuracy is achieved in practice and fast
algorithms were developed.  A new sampling theory on the ball was
developed by \cite{leistedt:flaglets} recently, augmenting the
sampling theorem on the sphere of \cite{mcewen:fssht} with Gaussian
quadrature on the radial line to recover an exact harmonic transform
suitable for the analysis of large data-sets defined on the ball.

The analysis of data defined on the sphere or ball often leads to data
defined on the rotation group \sothree, the space of three-dimensional
rotations.  Directional wavelet transforms on the sphere
(\eg\ \cite{antoine:1998, wiaux:2005, wiaux:2005c, wiaux:2005b,mcewen:2006:fcswt,
  mcewen:2006:cswt2, wiaux:2007:sdw, mcewen:2013:waveletsxv,
  mcewen:s2let_spin, chan:s2let_curvelets}) probe signal content not only in scale and
position on the sphere, but also in orientation.  The resulting
wavelet coefficients thus live on the rotation group.
Consequently, in all applications where directional spherical wavelet transforms are performed, 
%(\eg\ cosmology, biomedical imaging), 
sampling on the rotation group is important.
Moreover, the
wavelet transform can be computed via a Fourier transform on the
rotation group \cite{wiaux:2007:sdw, mcewen:2013:waveletsxv}.  Data
defined natively on the rotation group also arise in many
applications, for example searching databases of objects over
arbitrary rotations \cite{Garcia:2003}.  Sampling theorems on the
rotation group with fast harmonic transforms are thus of both
important theoretical interest and practical use.

The canonical equiangular sampling theorem on the rotation group
\sothree\ is that of Kostelec \etal\ \cite{kostelec:2008}, which
relies on the Driscoll \& Healy sampling theorem on the sphere
\cite{driscoll:1994}, and requires $\sim 8\elmax^3$ samples to
capture a signal on the rotation group that is band-limited at $\elmax$.
In this article we develop a novel sampling theorem on the rotation
group,
% (extending our recent sampling theorem on the sphere
%\cite{mcewen:fssht}), 
reducing the number of samples required to
capture a band-limited signal to $\sim 4\elmax^3$.  Furthermore, we
present fast algorithms to compute the associated harmonic transform
on the rotation group (often called the Wigner transform).  No existing sampling theorem on the rotation group reaches the optimal number of samples given by the $\sim 4\elmax^2/3$ degrees of freedom in harmonic space, although our approach comes closest to this bound. 

The remainder of this article is structured as follows.  After
concisely reviewing harmonic analysis on the rotation group we present
our novel sampling theorem, before defining explicitly an exact
quadrature rule on the rotation group.  We then discuss and evaluate
fast algorithms for computing the Wigner transform associated with our
sampling theorem, presenting a new public code, followed by concluding
remarks.
 
%==============================================================================
\section{Harmonic Analysis on the Rotation Group}

We consider the space of square integrable functions on the rotation
group $\ltwo(\sothree)$, with inner product  %of $f,g\in \ltwo(\sothree)$:
%\begin{equation*}
\mbox{$\innerp{f}{g} = \int_\sothree \deul{\eul} \: f(\eul) \: g^\cconj(\eul)$}
% \spcend ,
% \end{equation*}
for $f,g\in \ltwo(\sothree)$, where $\deul{\eul} = \sin\eulb \dx\eula
\dx\eulb \dx\eulc$ is the usual invariant measure on \sothree, which
is parameterised by the Euler angles $\eul=(\euls) \in \sothree$, with
$\eula \in [0,2\pi)$, $\eulb \in [0,\pi]$ and $\eulc \in [0,2\pi)$.
We adopt the $zyz$ Euler convention corresponding to the rotation of a
physical body in a \emph{fixed} coordinate system about the $z$, $y$
and $z$ axes by $\eulc$, $\eulb$ and $\eula$, respectively.

The Wigner $\dmatbig$-functions \Dlmn, with natural $\el\in\naturals$
and integer $\m,\n\in\integers$, $|\m|,|\n|\leq\el$, are the matrix
elements of the irreducible unitary representation of the rotation
group \sothree.  Consequently, the \Dlmnc\ also form an orthogonal
basis in $\ltwo(\sothree)$.\footnote{We adopt the conjugate
  \dmatbig-functions as basis elements since this convention
  simplifies connections to wavelet transforms on the sphere.}  The
orthogonality and completeness relations for the Wigner
$\dmatbig$-functions read, respectively,
%\begin{equation}
$
\innerp{\dmatbig_{\m\n}^{\el}}{\dmatbig_{\m\p\n\p}^{\el\p}}
= 
8\pi^2
\kron{\el}{\el\p}
\kron{\m}{\m\p}
\kron{\n}{\n\p} / (2\el+1)
$
%\end{equation}
and $ \sumlmn \dmatbig_{\m\n}^{\el}(\eula,\eulb,\eulc) \:
\dmatbig_{\m\n}^{\el\cconj}(\eula\p,\eulb\p,\eulc\p) =$ \mbox{$\delta(\eula -
\eula\p) \delta(\cos\eulb - \cos\eulb\p) \: \delta(\eulc - \eulc\p) $},
% \begin{align}
%   \sumlmn &
%   \dmatbig_{\m\n}^{\el}(\eula,\eulb,\eulc) \:
%   \dmatbig_{\m\n}^{\el\cconj}(\eula\p,\eulb\p,\eulc\p) \nonumber \\
%   & =
%   \delta(\eula - \eula\p)
%   \delta(\cos\eulb - \cos\eulb\p) \:
%   \delta(\eulc - \eulc\p)
%   \spcend ,
% \end{align}
where $\kron{i}{j}$ is the Kronecker delta symbol and
$\delta(x)$ is the Dirac delta function.
The Wigner functions may be decomposed as \cite{varshalovich:1989}
\begin{equation}
\label{eqn:d_decomp}
\dmatbig_{\m\n}^{\el}(\euls)
= {\rm e}^{-\img \m\eula} \:
\dmatsmall_{\m\n}^\el(\eulb) \:
{\rm e}^{-\img \n\eulc}
\spcend ,
\end{equation}
where the real polar $\dmatsmall$-functions can be computed by
recursion (\eg\ \cite{risbo:1996,trapani:2006}).  The Wigner
$\dmatsmall$-functions have the following Fourier series decomposition
\cite{nikiforov:1991}:
\begin{equation}
  \label{eqn:wigner_sum_reln}
  \dlmnb = \img^{\n-\m} \sum_{\m\p=-\el}^\el
  \dlmnhalfpi{\el}{\m\p}{\m} \:
  \dlmnhalfpi{\el}{\m\p}{\n} \:
  \exp{\img \m\p \eulb}
  \spcend ,
\end{equation}
where \mbox{$\dlmnhalfpi{\el}{\m}{\n} \equiv \dlmn (\pi/2)$}.  This
expression follows from a factoring of rotations as highlighted by
Risbo \cite{risbo:1996}.
We also note that the Wigner functions are related to the spin
spherical harmonics by \cite{goldberg:1967}
\begin{equation}
\label{eqn:ssh_wigner}
\sshfarg{\el}{\m}{\sas}{\spin} = (-1)^\spin 
\sqrt{\frac{2\el+1}{4\pi} } \:
\dmatbig_{\m,-\spin}^{\el\:\cconj}(\sab,\saa  ,0)
\spcend .
\end{equation}

A square integrable function defined on the rotation group may thus be
represented by its Fourier expansion
\begin{equation}
  \label{eqn:wig_inverse}
   \f(\eul) = 
   \sum_{\el=0}^{\infty} \frac{2\el+1}{8\pi^2} \sum_{\m=-\el}^{\el}
   \sum_{\n=-\el}^\el
   \wigc{\f}{\el}{\m}{\n} \Dlmnpc
   \spcend,
\end{equation}
where the Fourier coefficients are given by
\begin{equation}
  \label{eqn:wig_forward}
  \wigc{\f}{\el}{\m}{\n} = \innerp{\f}{\Dlmnc} 
  = \int_\sothree \deul{\eul} \: \f(\eul) \: \Dlmnp
  \spcend.
\end{equation}
The Fourier transform on the rotation group \sothree\ is often called
the Wigner transform,
% , where the forward transform is defined by
% \eqn{\ref{eqn:wig_forward}} and the inverse transform by
% \eqn{\ref{eqn:wig_inverse}}; 
while the Fourier coefficients $\wigc{\f}{\el}{\m}{\n}$ are often called
Wigner coefficients.

%==============================================================================
\section{Sampling Theorem}
\label{sec:sampling_theorem}

We develop a novel sampling theorem on the rotation group by making a connecting to the three-torus,\footnote{Although \sothree\ is not homeomorphic to the three-torus ${\rm T}^3$, by excluding a subset of measure zero from \sothree\ the resulting space can be embedded in ${\rm T}^3$, which is sufficient for sampling purposes.} extending similar
approaches on the sphere
\cite{mcewen:fssht,mcewen:fsht,huffenberger:2010} (following closely
the approach of our sampling theorem on the sphere
\cite{mcewen:fssht}).\footnote{Gauss-Legendre quadrature may also be
  used to define an efficient sampling theorem on the rotation group
  by a similar extension of the approach outlined in
  \cite{mcewen:fssht} and is developed in \cite{khalid:so3_gl}, where
  corresponding fast Wigner transforms are constructed by a separation
  of variables.  Although the asymptotic number of samples is $4\elmax^3$ in both cases, the approach described in this article requires fewer samples than Gauss-Legendre quadrature, which for small band-limits can be significant.}  Band-limited signals $\f \in \ltwo(\sothree)$ are
considered, with $\wigc{\f}{\el}{\m}{\n}=0$ for all $\el\geq\elmax$,
$\m\geq\mmax$ and $\n\geq\nmax$ (with $\mmax,\nmax\leq\elmax$). Our
sampling theorem is encapsulated in an exact computation of the Wigner
transform of \f\ from a finite set of samples.
%  (since the harmonic
% space of the rotation group is discrete).

We adopt an equiangular sampling of the rotation group with sample
positions given by 
\begin{equation}
  \label{eqn:nodes_alpha}
  \eulaiang = 
  \frac{2 \pi \eulai}{2\mmax-1}, 
  \quad \mbox{where } \eulai \in \{ 0,1,\dotsc,2\mmax-2 \}
  \spcend ,
\end{equation}
\begin{equation}
  \label{eqn:nodes_beta}
  \eulbiang = 
  \frac{\pi(2\eulbi+1)}{2\elmax-1}, 
  \quad \mbox{where } \eulbi \in \{ 0,1,\dotsc,\elmax-1 \}
  \spcend ,
\end{equation}
and
\begin{equation}
  \label{eqn:nodes_gamma}
  \eulciang = 
  \frac{2 \pi \eulci}{2\nmax-1}, 
  \quad \mbox{where } \eulci \in \{ 0,1,\dotsc,2\nmax-2 \}
  \spcend ,
\end{equation}
(see \cite{yershova:2010} for alternative samplings of $\sothree$).
To make the connection with the three-torus, the \eulb\ domain is
extended to $[0, 2\pi)$ by simply extending the domain of the \eulbi\
index to include $\{ \elmax,\elmax+1,\dotsc,2\elmax-1 \}$.  The number
of required samples is thus
$\bigl[(\elmax-1)(2\mmax-1)+1\bigr](2\nmax-1) \sim 4 \elmax \mmax
\nmax$, or $\sim 4\elmax^3$ samples when $\elmax = \mmax =
\nmax$.

By noting the Wigner decomposition of \eqn{\ref{eqn:d_decomp}} and the
Fourier series representation of \eqn{\ref{eqn:wigner_sum_reln}},
%that follows from a factoring of rotations \cite{risbo:1996}, 
the forward Wigner transform of \eqn{\ref{eqn:wig_forward}} may be
written
\begin{equation}
  \wigc{\f}{\el}{\m}{\n} 
  = \img^{\m-\n}
  \summptrunc 
  \dlmnhalfpim \:
  \dlmnhalfpin \:
  \Gmnm
  \spcend ,
\end{equation}
where
\begin{align}
  \Gmnm
  &= \int_\sothree \deul{\eul} \:
  \f(\eul) \:
  \exp{-\img(\m \eula + \m\p\eulb + \n \eulc)} \\
  &= \int_0^\pi \dx\eulb\sin\eulb \: \Gmnb \: \exp{-\img \m\p\eulb}
  \label{eqn:eulb_int}
  \spcend,
\end{align}
and
\begin{align}
  \Gmnb
  = & 
  \int_0^{2\pi} \dx \eula 
  \int_0^{2\pi} \dx \eulc
  \f(\eul) 
  \exp{-\img(\m \eula + \n \eulc)} \\
  = &(2\pi)^2 \sum_{\eulai=-(\mmax-1)}^{\mmax-1} 
  \sum_{\eulci=-(\nmax-1)}^{\nmax-1}
  \f(\eulaiang, \eulb, \eulciang) \nonumber \\
  & \times
  \exp{-\img(\m \eulaiang + \n \eulciang)}
  / \bigl[(2 \mmax - 1) (2 \nmax - 1) \bigr]
  \label{eqn:gmnb2}
  \spcend.
\end{align}
Since Wigner coefficients are not defined for $|\m|, |\n|>\el$, we
set them to zero to enforce the constraints $|\m|, |\n| \leq \el$
when the order of summations and integrals are interchanged.
The final expression of \eqn{\ref{eqn:gmnb2}} follows by appealing to
the %discrete and continuous 
orthogonality of the complex exponentials.

To recover a sampling theorem on the rotation group we
develop an exact quadrature for \eqn{\ref{eqn:eulb_int}} by extending $\Gmnb$ to the domain $[0,2\pi)$, through the
construction 
\begin{equation}
 \rGmnb = 
 \begin{cases}
   \: \Gmnb \: , & \eulb \in [0,\pi]\\
   \: (-1)^{\m+\n} \: \Gmn(2\pi - \eulb) \: , & \eulb \in (\pi,2\pi)
 \end{cases}
\end{equation}
(\cf\ \cite{mcewen:fssht,huffenberger:2010}), which  may  be
represented by its Fourier decomposition
\begin{equation}
 \label{eqn:gmnb_fourier}
 \rGmnb = 
 (2 \pi)^2 
 \sum_{\m\p=-(\elmax-1)}^{\elmax-1}
 \Fmnm \:
 \exp{\img \m\p \eulbiang}
 \spcend .
\end{equation}
Substituting \eqn{\ref{eqn:gmnb_fourier}} into
\eqn{\ref{eqn:eulb_int}}, one recovers the exact quadrature
\begin{align}
  \label{eqn:weight_conv}
  \Gmnm
  = 
  (2\pi)^2 
  \sum_{\m\p{}\p=-(\elmax-1)}^{\elmax-1}
  \Fmnmp \:
  \weight (\m\p{}\p - \m\p)
  \spcend,
\end{align}
where the weights are given by $ \weight(\m\p) = \int_0^\pi
\dx\eulb \sin\eulb \: \exp{\img \m\p \eulb}$, which can be evaluated
analytically \cite{mcewen:fssht}.
% \begin{align*}
%  \weight(\m\p)
%  &= \int_0^\pi \dx(\cos\eulb) \: \exp{\img \m\p \eulb}
%  = 
%  \begin{cases}
%    \: \pm \img \pi/2, & \m\p=\pm 1\\
%    \: 0, & \m\p \text{ odd},\:\m\p\neq\pm1\\
%    \: 2/(1-{\m\p}^2), & \m\p \text{ even}
%  \end{cases}
%  \spcend .
% \end{align*}

The inverse Wigner transform may be computed in an analogous manner,
again noting \eqn{\ref{eqn:d_decomp}} and 
\eqn{\ref{eqn:wigner_sum_reln}}.  The Fourier coefficients of
the extension of \f\ to the three-torus are first computed by
\begin{equation}
  \FImnm
  = \img^{\n-\m}
  \sum_{\ell=0}^{\elmax-1}
  \frac{2\ell+1}{8\pi^2} \:
  \dlmnhalfpim \:
  \dlmnhalfpin \:
  \wigc{\f}{\el}{\m}{\n} 
  \spcend ,
\end{equation}
followed by the three-dimensional Fourier transform
\begin{align}
  \f(\eulaiang, \eulbiang, \eulciang)
  = &
  \sum_{\m=-(\mmax-1)}^{\mmax-1}
  \sum_{\n=-(\nmax-1)}^{\nmax-1}
  \sum_{\m\p=-(\elmax-1)}^{\elmax-1}
  \FImnm\nonumber \\
  &\times \exp{\img(\m \eulaiang + \m\p\eulbiang + \n \eulciang)}
  \spcend .
\end{align}
The samples of \f\ computed over $\eulb \in (\pi, 2\pi)$ are discarded.

%==============================================================================
\section{Exact Quadrature}

Our sampling theorem on the rotation group can be used to define an
explicit quadrature rule for the integration of a band-limited
function.  Approximately a quarter of the number of samples of the
sampling theorem are required for the quadrature rule since
integrating a band-limited signal corresponds to computing
$\wigc{\f}{0}{0}{0}$ only (aliasing in higher order coefficients can
be tolerated).  The exact quadrature reads:
\begin{align}
  \label{eqn:quadrature}
  I 
  &= \int_\sothree \deul{\eul} \: \f(\eul) \\
  &= 
  \sum_{\eulai=0}^{\mmax-1} \:
  \sum_{\eulbi=0}^{\elmax-1} \:
  \sum_{\eulci=0}^{\nmax-1} \:
  \f(\eulaiang\p,\eulbiang, \eulciang\p) \: \qweight(\eulbiang)
  \spcend ,
\end{align}
where $\eulaiang\p = 2\pi\eulai/\mmax$ and
$\eulbiang\p = 2\pi\eulbi/\nmax$ (noting the reduced domain of
$\eulai$ and $\eulbi$).  The number of samples required is
\mbox{$\bigl[(\elmax-1)\mmax+1\bigr]\nmax \sim \elmax \mmax \nmax$}, or
$\sim \elmax^3$ samples when $\elmax = \mmax = \nmax$.  The quadrature
weights are defined by
\begin{equation}
  \qweight(\eulbiang) = 
  \frac{(2\pi)^2}{\mmax\nmax} 
  \Bigl[ 
  \weighttrans(\eulbiang) 
  + (1 - \kron{\eulbi,}{\elmax-1}) \: \weighttrans(\eulb_{2\elmax-2-\eulbi})
  \Bigr]
  \spcend ,
\end{equation}
where \cite{mcewen:fssht}
\begin{equation}
  \weighttrans(\eulbiang) =
  \frac{1}{2\elmax-1} \: \sum_{\m\p=-(\elmax-1)}^{\elmax-1} \:
  \weight(-\m\p) \: \exp{\img \m\p \eulbiang}
  \spcend .
\end{equation}

%==============================================================================
\section{Fast Algorithms}

Fast algorithms to compute the Wigner transforms associated with our
sampling theorem can be implemented through a separation of variables,
as described in \sectn{\ref{sec:sampling_theorem}}, using FFTs
throughout to compute Fourier transforms efficiently.  Furthermore,
\eqn{\ref{eqn:weight_conv}} can be computed more efficiently in
Fourier space, allowing aliasing in terms $|\m\p|>\elmax-1$ so that
only $\elmax$ samples are needed in $\eulb$ (as described in
\cite{mcewen:fssht}).  Rather than implement this algorithm from
scratch, however, we instead make the connection to spin spherical
harmonic transforms to leverage our existing
\sshtcode\footnote{\url{http://www.spinsht.org}} code
\cite{mcewen:fssht}.

The forward Wigner transform may be expressed as
\begin{equation}
  \label{eqn:wigner_forward_spherical_harmonic1}
  \f_\n(\eula,\eulb) = \frac{2\pi}{2\nmax-1} 
  \sum_{\eulci=-(\nmax-1)}^{\nmax-1} \:
  \f(\eula,\eulb, \eulciang) \:
  \exp{-\img \n \eulciang}
  \spcend ,
\end{equation}
where
\begin{align}
  \wigc{\f}{\el}{\m}{\n} =&
  (-1)^\n
  \sqrt{\frac{4\pi}{2\el+1}}
  \int_0^\pi \dx\eulb \sin\eulb
  \int_0^{2\pi} \dx \eula  \nonumber \\
  &\times 
  \f_\n(\eula,\eulb) \:
  \sshfargc{\el}{\m}{\eulb,\eula}{-\n}
  \label{eqn:wigner_forward_spherical_harmonic2}
  \spcend .
\end{align}
\Eqn{\ref{eqn:wigner_forward_spherical_harmonic2}} is simply a spin
spherical harmonic transform with spin number $-\n$, which may be
computed for each $\n$ with asymptotic complexity $\order(\elmax^3)$
by \sshtcode. The Fourier transform of
\eqn{\ref{eqn:wigner_forward_spherical_harmonic1}} can be computed by
an FFT in $\order(\elmax^2 \nmax \log_2 \nmax)$. The forward Wigner
transform can thus be computed with overall complexity
$\order(\nmax\elmax^3)$.

The inverse Wigner transform may be expressed as
\begin{align}
  \f_\n(\eula,\eulb) =& 
  (-1)^\n
  \sum_{\el=0}^{\elmax-1}
  \sum_{\m=-\min(\el,\mmax-1)}^{\min(\el,\mmax-1)}
  \sqrt{\frac{2\el+1}{16\pi^3}} \nonumber \\
  &\times
  \wigc{\f}{\el}{\m}{\n} \:
  \sshfarg{\el}{\m}{\eulb,\eula}{-\n}
  \spcend ,
  \label{eqn:wigner_inverse_spherical_harmonic1}
\end{align}
where
\begin{equation}
  \label{eqn:wigner_inverse_spherical_harmonic2}
  \f(\eula, \eulb, \eulciang)
  =
  \sum_{\n=-(\nmax-1)}^{\nmax-1} \:
  \f_\n(\eula,\eulb) \:
  \exp{\img \n \eulciang}
  \spcend .
\end{equation}
\Eqn{\ref{eqn:wigner_inverse_spherical_harmonic2}} is simply an
inverse spin spherical harmonic transform with spin number $-\n$,
which may be computed for each $\n$ with asymptotic complexity
$\order(\elmax^3)$ by \sshtcode.  The Fourier transform of
\eqn{\ref{eqn:wigner_inverse_spherical_harmonic2}} can be computed by
an FFT in $\order(\elmax^2 \nmax \log_2 \nmax)$.  The inverse Wigner
transform can thus be computed with overall complexity
$\order(\nmax\elmax^3)$.

For real functions on the rotation group we exploit the conjugate
symmetry relation $\wigc{\f}{\el\cconj}{\m}{\n} = (-1)^{\m+\n}
\wigc{\f}{\el}{(-\m)}{(-\n)}$ to reduce memory requirements and
computational time by a factor of two.

The fast Wigner transforms implementing our novel sampling theorem on
the rotation group are implemented in our new
\sothreecode\footnote{\url{http://www.sothree.org}} code, which uses
\sshtcode\ to compute spin spherical harmonics transforms and {\tt
  FFTW}\footnote{\url{http://www.fftw.org}} to compute Fourier
transforms.  The core code of \sothreecode\ is written in {\tt C}, while
{\tt Matlab} interfaces are also exposed.

%==============================================================================
\section{Evaluation}

\begin{figure}[t]
\centering
\subfigure[$\nmax=\elmax$]{\includegraphics[viewport=10 15 420 210,clip=true,width=\linewidth]{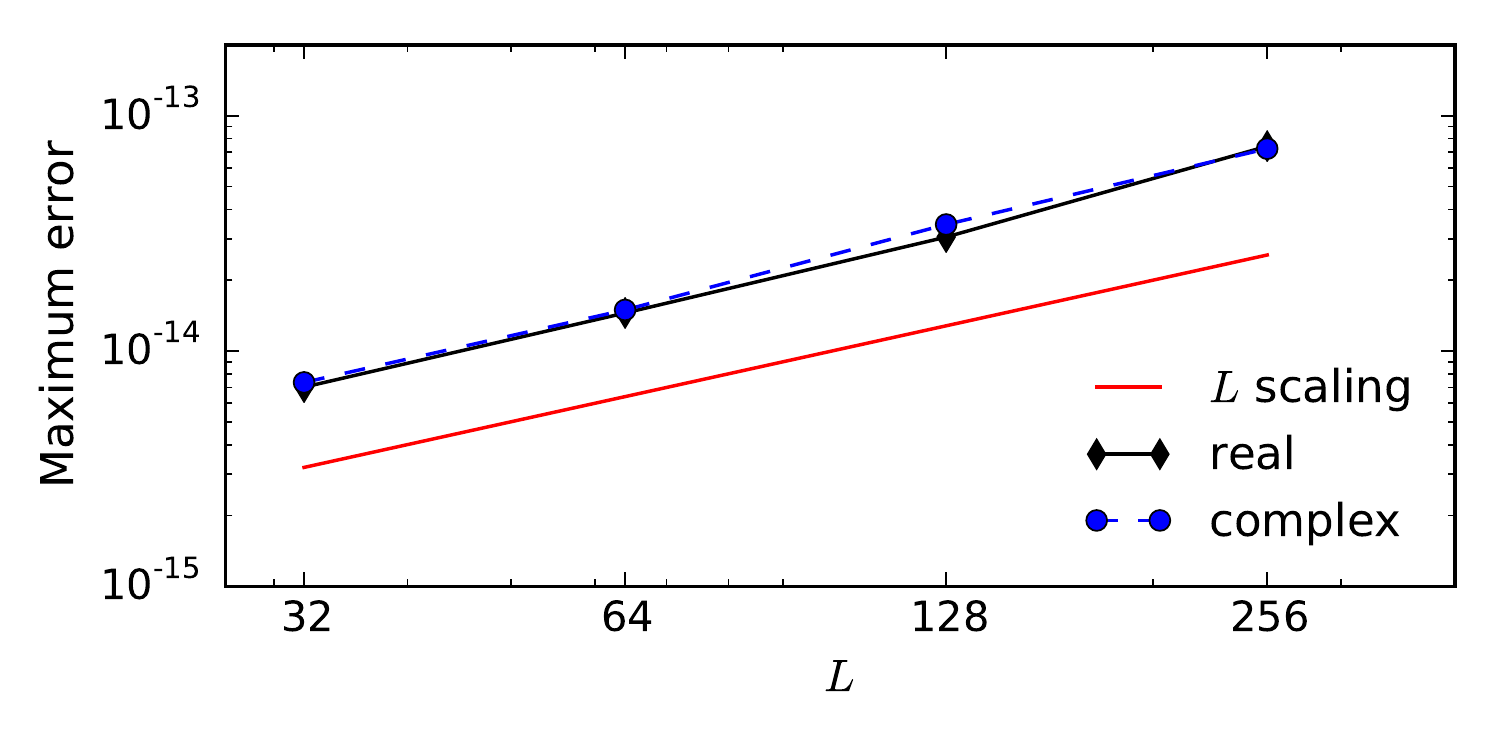}}
\subfigure[$\nmax=4$]{\includegraphics[viewport=10 15 420 210,clip=true,width=\linewidth]{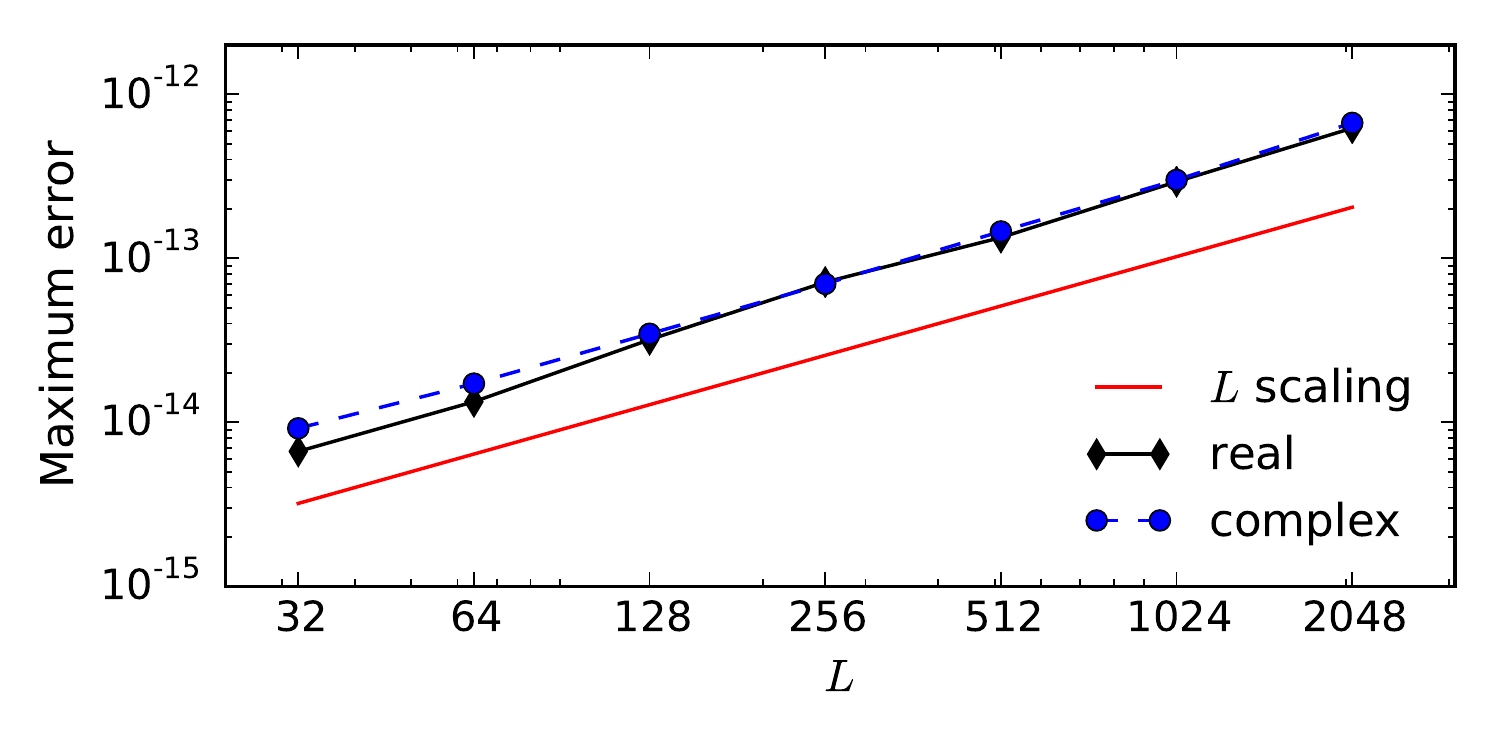}}
\caption{Numerical accuracy of the algorithms implementing our
  sampling theorem for real and complex signals for $\nmax=\elmax$
  (panel a) and $\nmax=4$ (panel b).  
  % The latter case is an example of
  % a common use-case, for example to compute directional wavelet
  % transforms of high-resolution (\ie\ high band-limit) data-sets
  % defined on the sphere using wavelets with low azimuthal
  % band-limits. 
  Accuracy is found empirically to scale approximately
  linearly with increasing band-limit.}
\label{fig:accuracy}
\end{figure}

\begin{figure}[t]
\centering
\subfigure[$\nmax=\elmax$]{\includegraphics[viewport=10 15 420 210,clip=true,width=\linewidth]{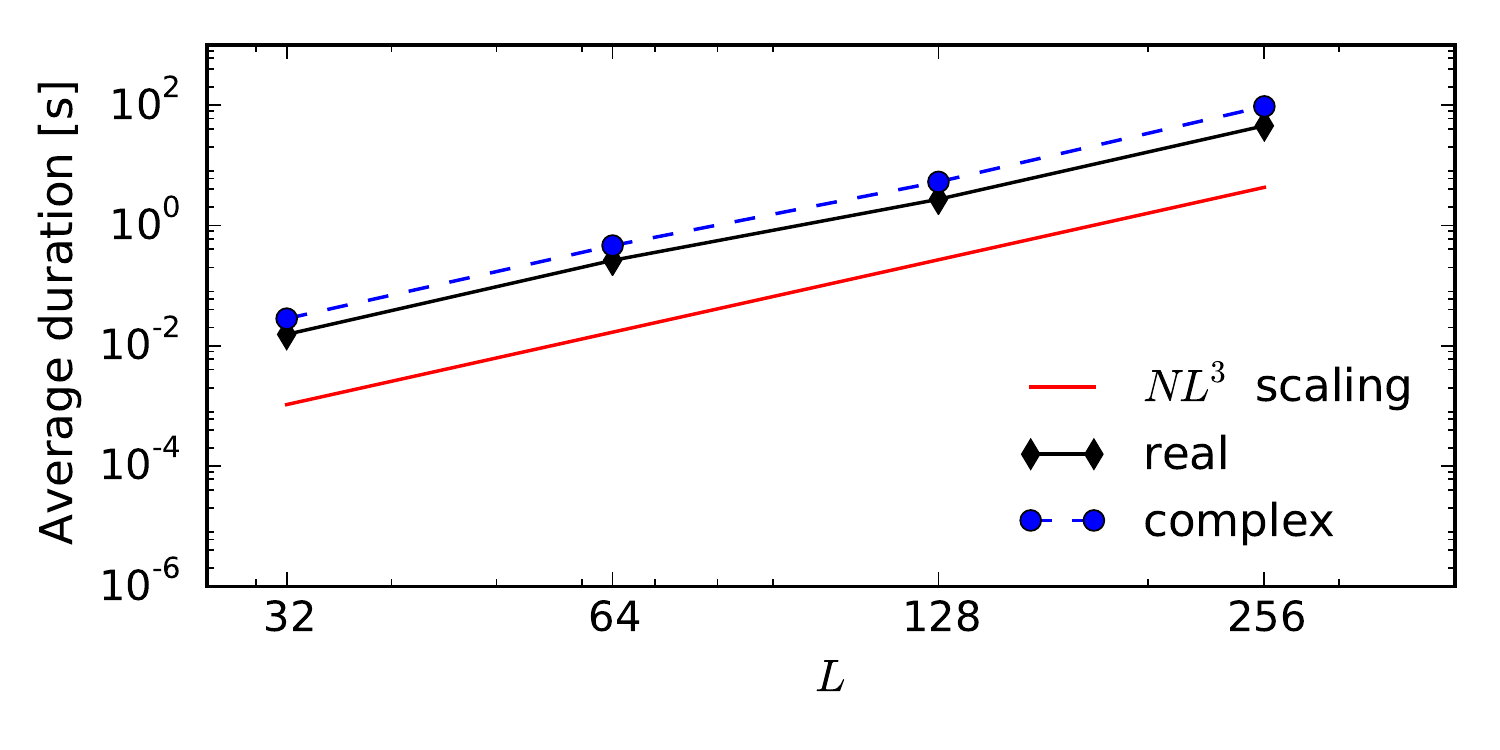}}
\subfigure[$\nmax=4$]{\includegraphics[viewport=10 15 420 210,clip=true,width=\linewidth]{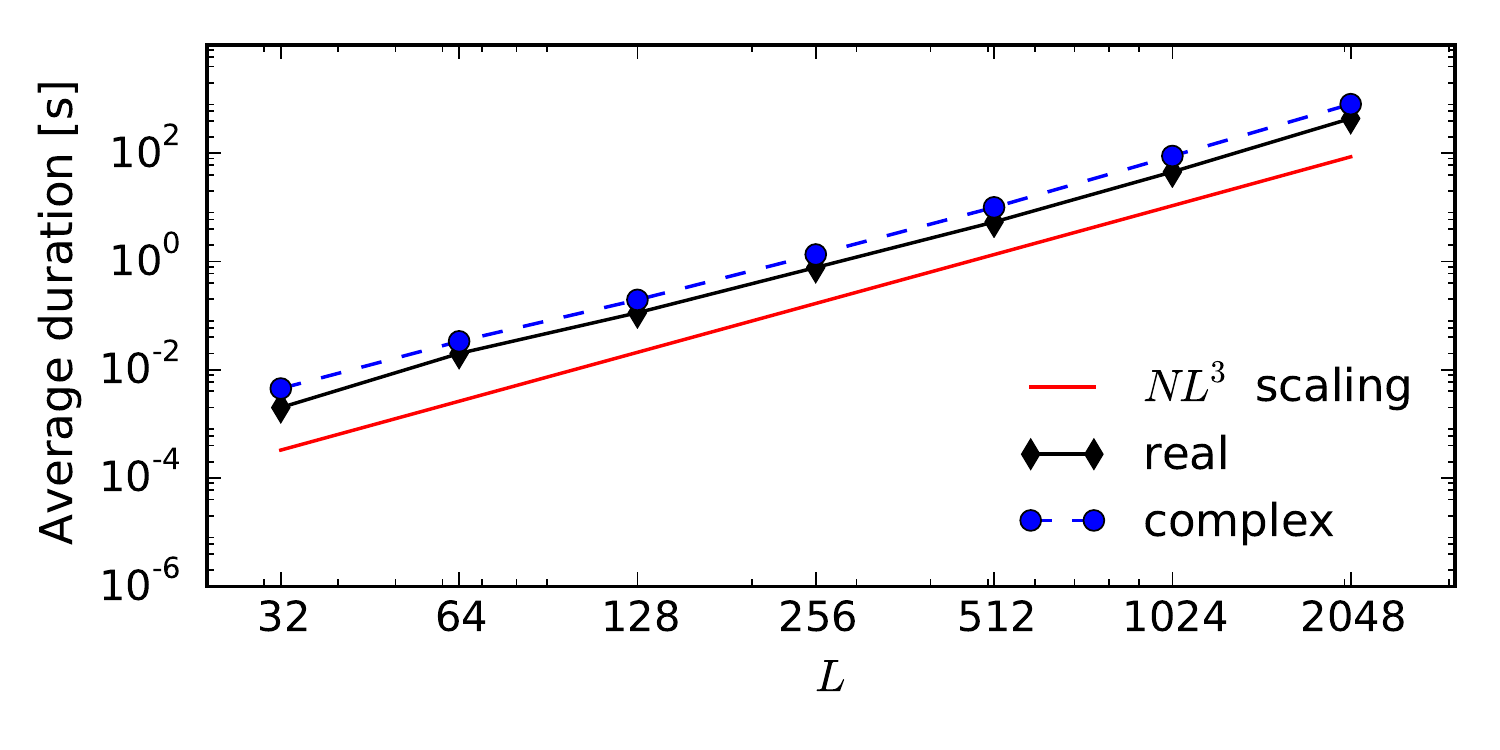}}
\caption{Computation time of the algorithms implementing our sampling
  theorem for real and complex signals for $\nmax=\elmax$ (panel a)
  and $\nmax=4$ (panel b).  
  % The latter case is an example of a common
  % use-case, for example to compute directional wavelet transforms of
  % high-resolution (\ie\ high band-limit) data-sets defined on the
  % sphere using wavelets with low azimuthal band-limits. 
  As predicted, computation time scales as $\order(\nmax \elmax^3)$
  and real transforms are approximately twice as fast as complex
  transforms (since additional symmetries are exploited).}
\label{fig:timing}
\end{figure}

In order to assess the numerical accuracy and computation time of our
algorithms we perform the following numerical experiment.  We generate
band-limited test signals defined by uniformly random
Wigner coefficients with real and imaginary parts distributed in the
interval $[-1,1]$.  An inverse transform is performed to synthesise
the test signal on the rotation group from its Wigner coefficients,
followed by a forward transform to recompute Wigner coefficients.
Numerical accuracy is measured by the maximum absolute error between
the original Wigner coefficients and the recomputed values.
Computation time is measured by the average of the times taken to
perform the inverse and forward transform.   

All numerical experiments are performed on a single core of a 2.67 GHz
Intel(R) Xeon(R) CPU X5650 machine with 100 GB of RAM and are averaged over ten random
test signals.  We consider two modes of operation: (i) $\nmax=\elmax$
and (ii) a small constant \nmax, in this example $\nmax=4$.  The
latter mode of operation is a common use-case, for example when
computing directional wavelet transforms on the sphere for wavelets
with low azimuthal band-limit, and allows one to consider very large
harmonic band-limits \elmax.  Note that the results presented here are
obtained using version 1.1b1 of \sshtcode, which contains many
computational optimisations compared to previous versions.  The Risbo
recursion \cite{risbo:1996} for computing Wigner
$\dmatsmall$-functions is adopted.
%  here, although we note that for
% band-limits up to $\elmax=2048$ the recursion of Trapani \& Navaza
% \cite{trapani:2006} can be used to reduce computation time (see \eg\
% \cite{mcewen:fssht}).

The maximum absolute error is plotted against the band-limit in
\fig{\ref{fig:accuracy}}.  High numerical accuracy is achieved, with
errors on the order of machine precision and found empirically to
increase approximately linearly with band-limit.

The computation time for both real and complex signals on the rotation
group is plotted against the band-limit in \fig{\ref{fig:timing}}.
Computation time evolves as $\order(\nmax\elmax^3)$ as
predicted.  Computation time for real signals is approximately twice as
fast as for complex signals, also as predicted.

%==============================================================================
\section{Conclusions}

We have presented a new sampling theorem on the rotation group
\sothree\ and fast algorithms to compute the associated Wigner
transform.  Our sampling theorem requires $4\elmax^3$ samples to
capture the full information content of a signal band-limited at
$\elmax$, reducing the number of required samples by a factor of two
compared to other equiangular sampling theorems on the rotation group
\cite{kostelec:2008}.  Our fast algorithms are theoretically exact and
achieve accuracy close to machine precision, while scaling as
$\order(\elmax^4)$, compared to the naive scaling of
$\order(\elmax^6)$.  For the common case of a low directional
band-limit $\nmax$ the scaling is reduced to $\order(\nmax\elmax^3)$.
In a separate article \cite{mcewen:s2let_spin} we apply these fast
algorithms to improve the efficiency of a directional wavelet
transform on the sphere \cite{wiaux:2007:sdw, mcewen:2013:waveletsxv}
in order to support the analysis of large data-sets.

\newpage

\ifCLASSOPTIONcaptionsoff
  \newpage
\fi

%\IEEEtriggeratref{16}


%\bibliographystyle{IEEEtran}

\bibliographystyle{IEEEtran}
\bibliography{bib_myname,bib_journal_names_long,bib}



% \end{thebibliography}


% \end{IEEEbiography}


% % insert where needed to balance the two columns on the last page with
% %\newpage



%\vfill


\end{document}